# Aerocapture Enabled Fast Uranus Orbiter Missions

Athul Pradeepkumar Girija [1**][**]

[1]*School of Aeronautics and Astronautics, Purdue University, West Lafayette, IN 47907, USA*

## ABSTRACT

At the far reaches of the outer Solar System, the ice giants remain the last class of planets yet to be studied using orbiters. The 2023-2032 Planetary Science Decadal Survey has underscored the importance of the ice giants in understanding the origin, formation, and evolution of our Solar System. The enormous heliocentric distance of Uranus presents considerable mission design challenges, the most important being able to reach Uranus within a reasonable time. The present study presents two examples of aerocapture enabled short flight time, fast trajectories for Uranus orbiter missions, and highlights the enormous benefits provided by aerocapture. The first is an EEJU trajectory with a launch opportunity in July 2031 with a flight time of 8 years. The second is an EJU trajectory with a launch opportunity in June 2034 with a flight time of only 5 years. Using the Falcon Heavy Expendable, the available launch capability is 4950 kg and 1400 kg respectively for the two trajectories. Both trajectories have a high arrival speed of 20 km/s, which provides sufficient corridor width for aerocapture. Compared to propulsive insertion architectures which take 13 to 15 years, the fast trajectories offer significant reduction in the flight time.

***Keywords:*** Uranus orbiter, Fast missions, Gravity-assist, Aerocapture

---

[****] To whom correspondence should be addressed, E-mail: apradee@purdue.edu

## I. INTRODUCTION

At the far reaches of the outer Solar System, the ice giants Uranus and Neptune remain the last class of planets which are yet to be studied using orbiters [1, 2]. The fleeting encounter by Voyager 2 nearly 40 years ago revealed Uranus to be a tantalizing world with many mysteries hidden underneath its distinctive blue-green, but rather featureless atmosphere at the time [3]. The 2023-2032 Planetary Science Decadal Survey has underscored the importance of the ice giants in understanding the origin, formation, and evolution of our Solar System. A Flagship-class mission to Uranus is recommended as the top priority for a large strategic mission in the next decade [4]. The enormous heliocentric distance of Uranus (19 AU) presents considerable mission design challenges, the most important being able to reach Uranus within a reasonable time after launch [5]. For conventional propulsive insertion, the other major challenge is to minimize the approach speed at Uranus, so as to minimize the orbit insertion $\Delta V$ [6]. Balancing the two competing requirements is a major challenge, as a shorter time of flight mission will generally drive up the approach speed, increasing the orbit insertion $\Delta V$, and the required propellant mass [7, 8]. A promising alternative to propulsive insertion that overcomes this challenge is the technique of aerocapture (Figure 1), which uses atmospheric drag from a single pass to achieve nearly propellant-free orbit insertion even when the approach speeds are high [9, 10]. The present study presents two examples of aerocapture enabled short flight time, fast trajectories for Uranus orbiter missions, and highlights the enormous benefits provided by aerocapture for future ice giant missions.

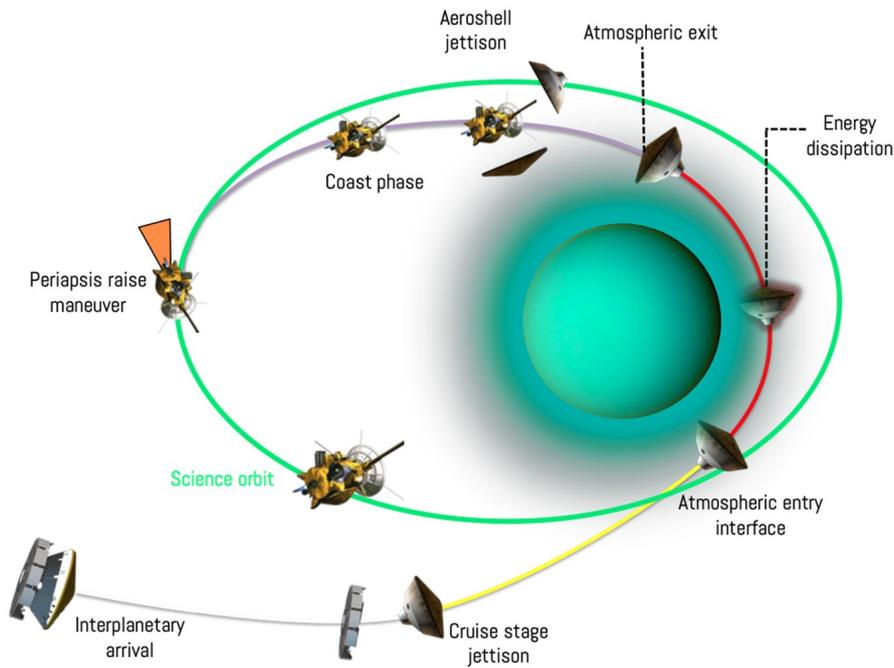

Figure 1. Illustration of the aerocapture maneuver at Uranus using an MSL-derived aeroshell.

## II. MISSION DESIGN TRADE SPACE

Figure 2 shows a high-level overview of the available interplanetary trajectory options to Uranus in the next decade [11]. The top level mission constraints are $C_3 <= 120$, and arrival v_inf <= 22 km/s for the gravity-assist paths shown in the figure. The objective of the study is to identify viable short flight time trajectories for orbit insertion with aerocapture. The left panel shows the trade-off between launch $C_3$ and the time of flight. The right panel shows the trade-off between the arrival v_inf and the time of flight. The launch $C_3$ is an important parameter because it dictates the amount of mass that can be launched on a given launch vehicle. Higher the $C_3$, less mass can be launched. If the $C_3$ is too high, it may render the trajectory infeasible with certain launchers [12]. For the present study, the highest capability launcher assumed to be available is the Falcon Heavy Expendable with a STAR-48 upper stage. Trajectories near the bottom right corner of the right panel (ToF = 13–15 years, Vinf = 6–8 km/s) are generally considered for propulsive insertion architectures. The proposed baseline Uranus Orbiter and Probe (UOP) Flagship mission, for example uses a 13-year time of flight trajectory with an arrival speed of 6.3 km/s [13]. Previous studies have shown that on the contrary, aerocapture with lift modulation control requires high arrival speeds of 20 km/s to provide a sufficiently large corridor ~ 1 deg [14, 15]. From Figure 2, it is seen that there is a set of EJU ($C_3$ = 120) and EEJU ($C_3$ = 55) trajectories which satisfy the approach speed requirement for aerocapture (~20 km/s) while allowing flight times as short as 5 and 8 years respectively. These trajectories are entirely infeasible with propulsive insertion, as the orbit insertion ΔV exceeds 9 km/s (550,00 km apoapsis) which is prohibitive. However, aerocapture essentially offers unlimited ΔV, making it an enabling orbit insertion technology for these fast arrival trajectories [16].

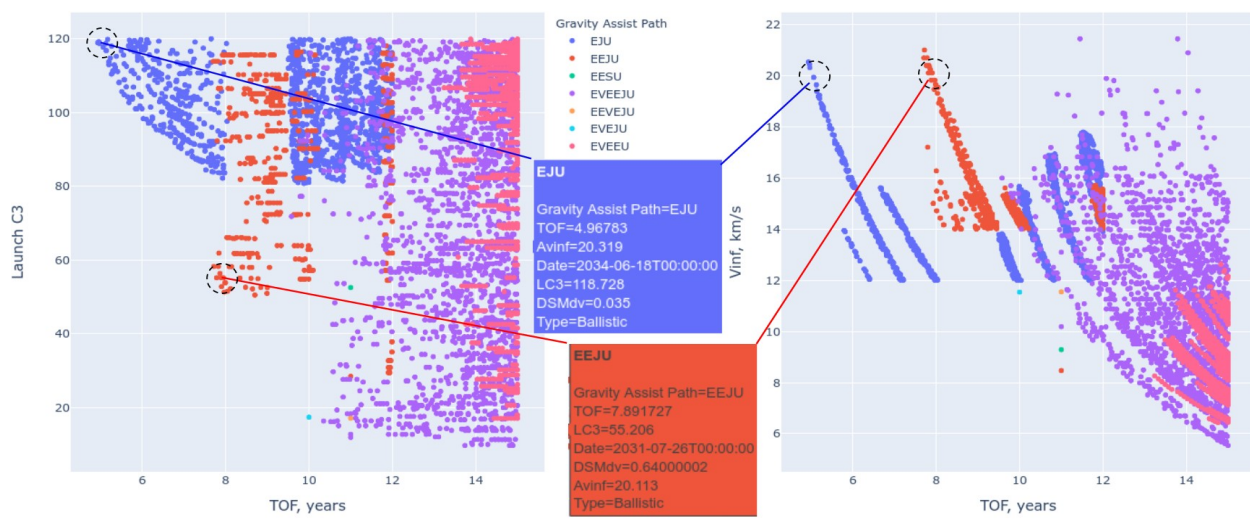

Figure 2. Trajectory trade space for Uranus, with two fast arrival options highlighted.

Figure 3 shows the available launch mass capability with the FH Expendable for these trajectories. For the lower $C_3$ E(ΔV)EJU launch opportunity in July 2031 ($C_3$=55, ToF=8y), the launch capability is 4950 kg. This has been shown to be sufficient for a 1400 kg Uranus orbiter and a 300 kg atmospheric entry probe with an MSL-derived aeroshell [17]. For the higher C3, EJU opportunity in June 2034 (C3=120, ToF=5y), the launch capability is 1400 kg which allows for an approximately 500 kg orbiter, but no atmospheric probe. It is worth noting that the flight times of these trajectories are considerably shorter than that for baseline propulsive architectures which generally take 13–15 years. Figure 4 shows the approach trajectory to Uranus atmospheric entry interface for aerocapture into a polar orbit.

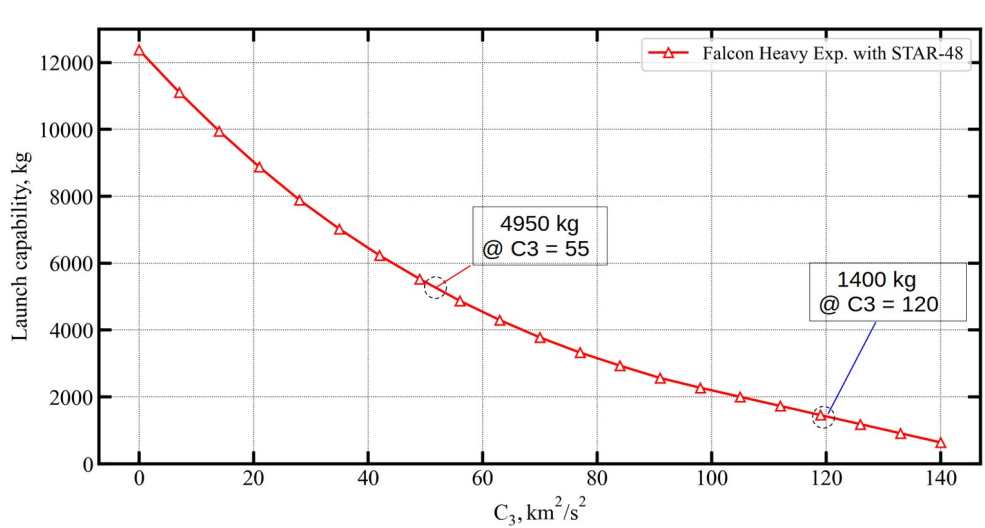

Figure 3. Falcon Heavy Expendable with STAR-48 high energy launch performance curve.

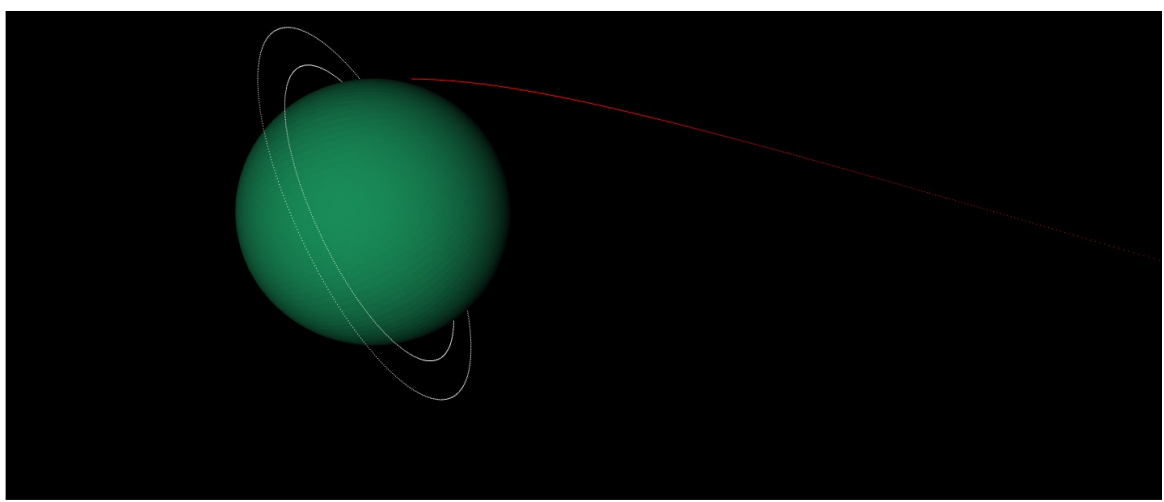

**Approach Results (Entry state / orbital parameters)**

| Altitude, km | Longitude, deg (BI) | Latitude, deg (BI) | Speed, km/s (atm. rel) | EFPA, deg. (atm. rel) | Heading angle, deg. |
|---|---|---|---|---|---|
| 1000 | -13.98 | 36.4 | 29.39 | -11.58 | 84.98 |

Figure 4. Approach trajectory to Uranus atmospheric entry interface for aerocapture.

## III. AEROCAPTURE TRAJECTORY

Aerocapture at Uranus has been studied using both lift and drag modulation control techniques [18, 19]. Lift modulation is chosen for this study because it is better suited for the high speed arrival trajectories in terms of corridor width and aero-thermal environments. The vehicle hits the atmosphere at about 29.3 km/s, for which the aerocapture corridor bounds are approximately [-12.00, -11.00] deg. with a width of 1 deg. This allows for about ±0.3 deg entry-flight path angle error (3σ), and provides some margin for atmospheric uncertainties which can be quite large at Uranus. Figure 5 shows the nominal undershoot (steep) and overshoot (shallow) limit aerocapture trajectories using the MSL-derived aeroshell. The lowest altitude attained during aerocapture is about 300 km above the 1 bar pressure-level, during which the speed drops from 29 km/s (planet-relative) to about 20 km/s. The atmospheric drag provides about 9 km/s of ΔV during the maneuver to target a 550,000 km apoapsis. The peak deceleration is in the range of 4–10 g, which is well within the range of decelerations typical entry vehicles are subjected to [20, 21]. The peak heat rate is in the range of 1400–1800 W/cm$^2$, which is well within the tested limits of the HEEET thermal protection system [22], and much less than that encountered during the steep entry of an entry probe [23]. The integrated heat load is about 200–300 kJ/cm$^2$, which is quite high but still reasonable for HEEET. Even with the high heat load, the TPS mass fraction is expected to be < 25% due to the advanced design of HEEET [24]. Based on historical data, the structural mass fraction is also estimated at 25%, leaving about 50% of the vehicle entry mass for the orbiter and the probe.

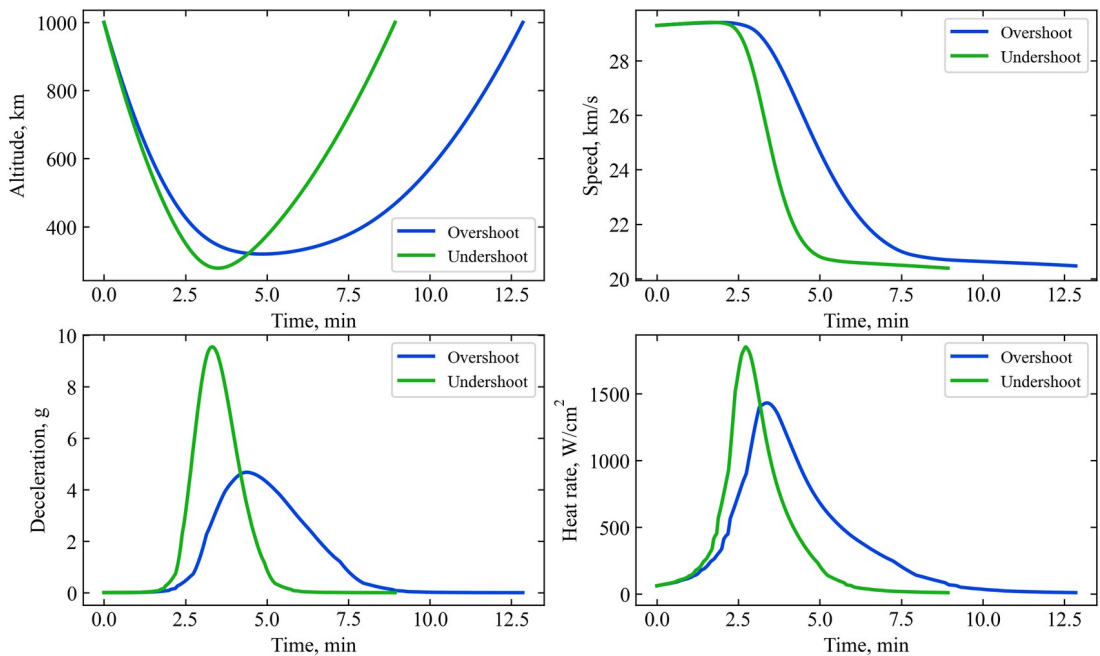

Figure 5. Nominal undershoot and overshoot aerocapture trajectories with L/D = 0.24.

After aerocapture, the spacecraft performs an 80 m/s periapsis raise maneuver to enter its initial 4,000 x 500,000 km capture orbit. Figure 6 shows the approach trajectory and the initial capture orbit achieved after aerocapture.

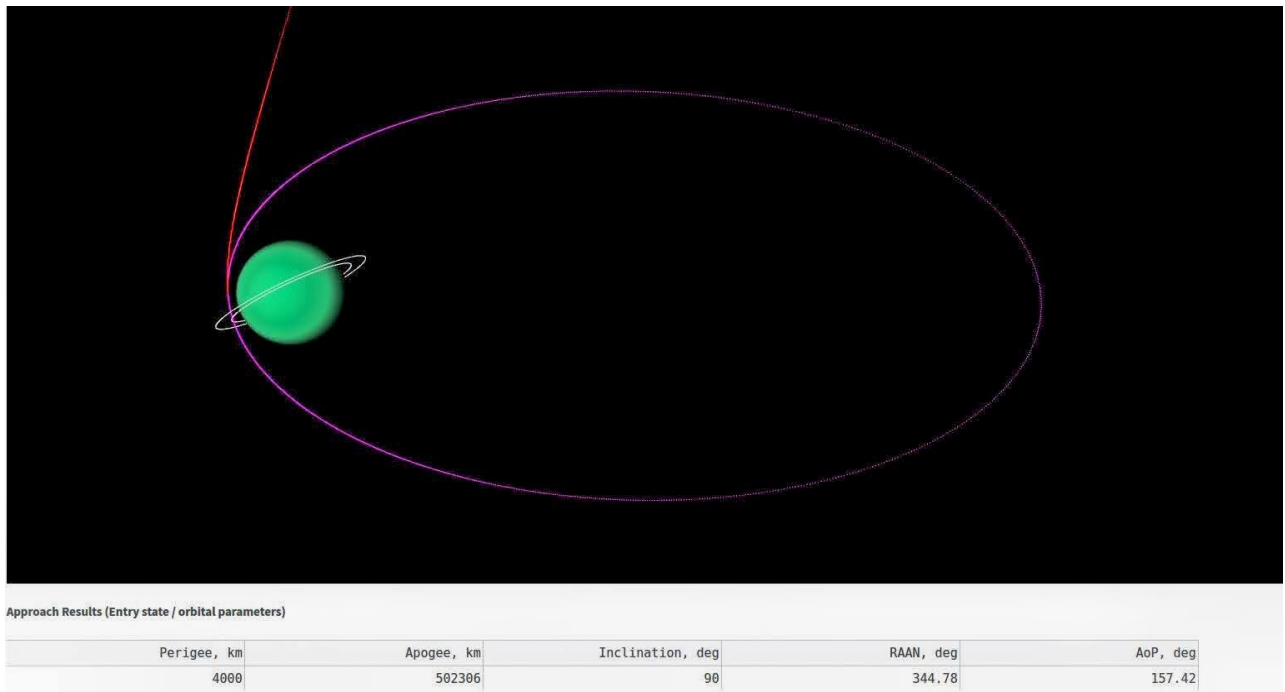

Approach Results (Entry state / orbital parameters)

| Perigee, km | Apogee, km | Inclination, deg | RAAN, deg | AoP, deg |
|---|---|---|---|---|
| 4000 | 502306 | 90 | 344.78 | 157.42 |

Figure 6. The spacecraft in its 4000 x 500,000 km polar orbit at Uranus after aerocapture.

## IV. IMPLICATIONS FOR FUTURE ICE GIANT MISSIONS

Aerocapture has been studied for over six decades for its applications to Solar System exploration from Venus to Neptune [25, 26]. Most aerocapture studies have almost exclusively been for Mars and Venus missions, with some studies for Titan [27, 28]. Within the last decade, considerable studies have shown the great potential offered by drag modulation for small satellite missions to Mars and Venus [29, 30]. For example, aerocapture has been shown to enable small low-cost standalone missions to Venus [31], small missions as elements of New Frontiers of Flagship missions [32], and atmospheric sample return from the Venusian cloud layers [33]. Aerocapture has also recently been proposed Titan orbiter missions within New Frontiers [34]. However, the benefits offered by aerocapture for ice giant missions is the most significant of all Solar System destinations, by enabling the use of short flight time, fast arrival trajectories which are simply impractical with propulsive insertion [35]. The technical risk posed by navigation and atmospheric uncertainties is unfortunately, also the greatest at the ice giants [36]. Until in-situ data from a probe becomes available, ice giant aerocapture guidance schemes need to be able to demonstrate a very high degree of robustness against the large atmospheric uncertainties [37]. Even though considerable challenges face a future mission, the benefits offered by aerocapture are substantial in enabling fast missions to these distant worlds [38, 39].

## V. CONCLUSIONS

The study presented two examples of aerocapture enabled short flight time, high arrival speed trajectories for future Uranus orbiter missions. The first is an EEJU trajectory with a launch opportunity in July 2031 with a $C_3 = 55$, and flight time of 8 years. The second is an EJU trajectory with a launch opportunity in June 2034 with a $C_3 = 120$, and a flight time of 5 years. Both trajectories have a high arrival speed of 20 km/s, which provides sufficient corridor width for an MSL-derived aeroshell with L/D = 0.24. The aerocapture maneuver provides a ΔV of about 9 km/s which is so high that propulsive mission studies do not even consider such interplanetary trajectory options. Compared with baseline propulsive insertion architectures which take 13–15 years to reach Uranus, these aerocapture enabled fast arrival trajectories offer significant reduction in the flight time. The estimated aero-thermal loads are well within the capability of HEEET thermal protection system which has been designed to withstand even more severe conditions for ice giant probe entry. Due to their extreme heliocentric distance, the benefits offered by aerocapture for ice giants is the most significant of all Solar System destinations. By enabling orbit insertion from short flight time fast arrival trajectories, aerocapture is a mission enabling technology for fast missions to explore these distant worlds.

## DATA AVAILABILITY

The aerocapture trajectory results can be reproduced using the Aerocapture Mission Analysis Tool (AMAT) v2.2.22. Some of the data and the code used to make the study results will be made available upon reasonable request.